# Negative differential resistance in Josephson junctions coupled to a cavity

N.F. Pedersen [a], G. Filatrella [b,*], V. Pierro [c], M.P. Sørensen [a]

[a] Department of Applied Mathematics and Computer Science, Technical University of Denmark, 2800 Kgs. Lyngby, Denmark
[b] Department of Science and Technology, University of Sannio, Italy
[c] Department of Engineering, University of Sannio, Italy

ABSTRACT

Regions with negative differential resistance can arise in the IV curve of Josephson junctions and this phenomenon plays an essential role for applications, in particular for THz radiation emission. For the measurement of high frequency radiation from Josephson junctions, a cavity – either internal or external – is often used. A cavity may also induce a negative differential resistance region at the lower side of the resonance frequency. We investigate the dynamics of Josephson junctions with a negative differential resistance in the quasi particle tunnel current, i.e. in the McCumber curve. We find that very complicated and unexpected interactions take place. This may be useful for the interpretation of experimental measurements of THz radiation from intrinsic Josephson junctions.

## 1. Introduction

Stacks of BSCCO consisting of coupled Josephson Junctions (JJ) can give rise to interesting emission in the THz region [1–4]. The technological applications of THz electromagnetic radiation span in the field of security, nondestructive sensing, biological sensors, fast communications, and analogical processing [5]. Several technologies for THz management have been proposed. However, such frequency range is difficult to access, because it belongs to the so-called "terahertz gap".

The interaction between JJ and the cavity is therefore an essential step for applications, inasmuch as the power emitted by the JJ is coupled to an external load that, at a simple level, can be modeled by an *RLC* circuit [6]. For instance, a characterization of the radiation power emitted by an antenna has been proposed in Ref. [7], where it has been shown that the antenna is well described by an *RLC* circuit.

JJ emission is associated with Negative Differential Resistance (NDR) in the *RLC* circuit together with the nonlinear Josephson junction [8,9]. In this work we will examine the interplay between the coupling circuit *RLC* and the JJ with NDR in the resistive part of the IV curve. Finally, we mention that arrays of JJ [10,11] are important for applications, as a single junction does not provide enough power for many practical purposes. The NDR considered in this work is a phenomenological approach to the more complicated dynamics arising in JJ arrays interaction [12,13].

## 2. Model

The resistively shunted Josephson junction (RSJ) equivalent circuit coupled to a cavity is schematically depicted in Fig. 1a [6]. The phase difference across the Josephson barrier is denoted $\phi = \phi(t)$ and the voltage across the junction is given by $V(t) = \hbar/(2e) d\phi/dt$, where $\hbar$ is Planck's constant divided by $2\pi$ and $e$ is the electron charge. The Josephson critical current is $I_0$, $C_J$ is the junction capacitance and $R_J$ denotes the quasiparticle tunneling resistance. The bias current is $I$. For the external linear cavity coupled to the Josephson junction circuit, the charge on the cavity capacitor is $\tilde{q} = \tilde{q}(t)$ and the cavity capacitance is $C$. In the cavity circuit we have inserted a resistor with resistance $R$ and inductor with inductance $L$. For a Josephson junction with a linear cavity and linear shunt resistance (McCumber resistance) we get [7,14,15]:

$$\frac{C_J\hbar}{2e}\frac{d^2\phi}{dt^2} + \frac{\hbar}{R_J 2e}\frac{d\phi}{dt} + I_0 \sin\phi + \frac{d\tilde{q}}{dt} = I \quad (1)$$

$$\frac{d^2\tilde{q}}{dt^2} + \frac{R}{L}\frac{d\tilde{q}}{dt} + \frac{1}{LC}\tilde{q} - \frac{\hbar}{2eL}\frac{d\phi}{dt} = 0. \quad (2)$$

Introducing the Josephson plasma frequency $\omega_0 = \sqrt{2eI_0/C_J\hbar}$, Eq. (1) can be cast in normalized units $\tau = \omega_0 t$ and $q = \omega_0\tilde{q}/I_0$:

* Corresponding author. Tel.: +39 3294173480.
*E-mail address:* filatrella@unisannio.it (G. Filatrella).

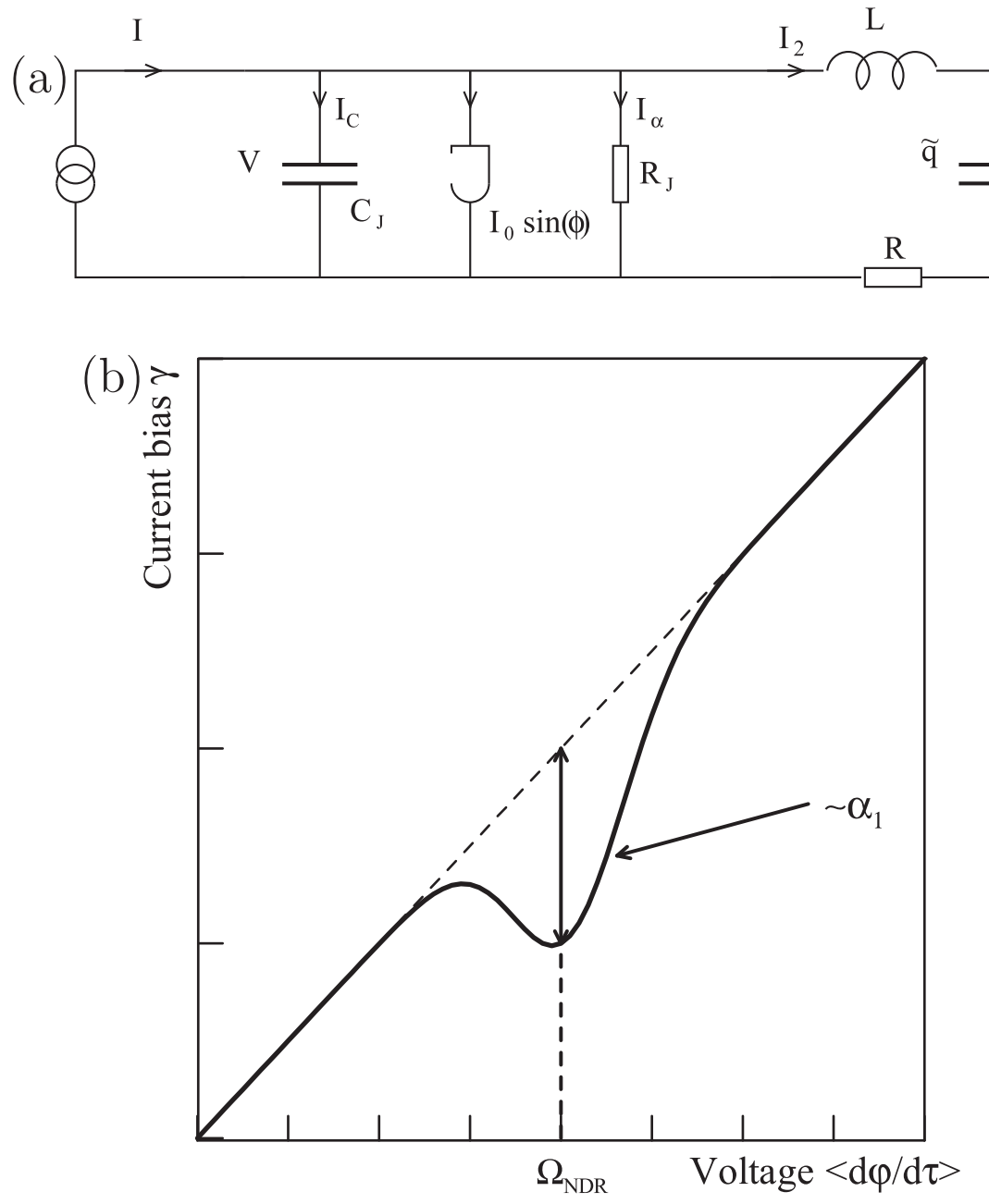

**Fig. 1.** (a) Equivalent circuit of a JJ coupled to a cavity. (b) The IV curve of the nonlinear conductance, Eq. (5). The dashed line is a linear conductance ($\alpha_1 = 0$).

$$\frac{d^2\phi}{d\tau^2} + \alpha\frac{d\phi}{d\tau} + \sin\phi + \frac{dq}{d\tau} = \gamma, \tag{3}$$

$$\frac{d^2q}{d\tau^2} + \frac{\Omega_r}{Q}\frac{dq}{d\tau} + \Omega_r^2 q - \beta_L\frac{d\phi}{d\tau} = 0, \tag{4}$$

where

$$\frac{Q}{\Omega_r} = \frac{L}{R}\sqrt{\frac{2eI_0}{C_J\hbar}},\ \beta_L = \frac{2eLI_0}{\hbar},\ \gamma = \frac{I}{I_0},\ \Omega_r = \frac{1}{\omega_0\sqrt{LC}}.$$

This model can be modified to account for a nonlinear conductance observed in experiments and in simulations of stacks of Josephson junctions [12,13]. A possibility is to introduce the nonlinear dissipation coefficient [16] in normalized units as follows

$$\alpha(d\phi/d\tau) = \alpha_0\left[1 - \alpha_1\exp\left(-\frac{1}{\Delta}\left(\frac{d\phi}{d\tau} - \Omega_{NDR}\right)^2\right)\right], \tag{5}$$

that models the effect of the complicated interaction in the BSCCO stack. Here $\alpha_0 = \frac{1}{R_J}\sqrt{\frac{\hbar}{2eI_0C_J}}$, is the normalized linear quasi particle conductance and $\alpha_1$ is a measure of the height of the Gaussian, see Fig. 1. The parameter $\Omega_{NDR}$ determines the normalized voltage where $\alpha$ takes its minimum value, while the parameter $\Delta$ is connected to the width of the NDR region. Examples of a nonlinear McCumber curve could be at the energy gap in intrinsic Josephson junctions and at the gap difference in junctions made of different materials.

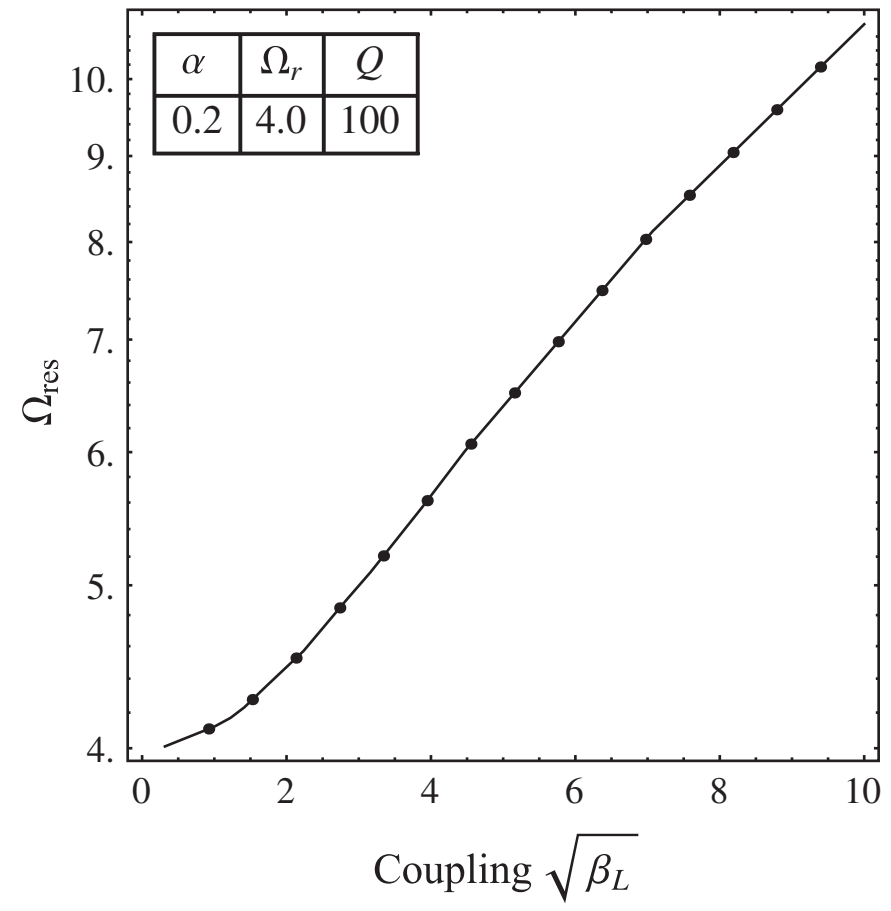

**Fig. 3.** The resonant frequency $\Omega_{res}$ as a function of the square root of the coupling $\sqrt{\beta_L}$.

The modified conductance changes the Josephson voltage relation and Eq. (4) becomes

$$\frac{d^2\phi}{d\tau^2} + \alpha\left(\frac{d\phi}{d\tau}\right)\frac{d\phi}{d\tau} + \sin(\phi) = -\frac{d\tilde{q}}{d\tau} + \gamma_B. \tag{6}$$

That reduces to the linear case for $\alpha_1 = 0$.

## 3. Results

To understand the combined effect of a JJ and an *RLC* we start simulating the IV curve and the current oscillations in the *RLC* circuit. To estimate the power available in the cavity we introduce the quantity.

$$A = \max_\tau \frac{dq}{d\tau} - \min_\tau \frac{dq}{d\tau} \tag{7}$$

We start with the NDR $\alpha_1 = 0.5$. An example is shown in Fig. 2. From the figure it is evident that the resonance of the *RLC* circuit is shifted to higher values, $V \simeq 5$ with respect to the resonance frequency $1/\omega_0\sqrt{LC} = \Omega_r = 4$ of the resonator. This is understandable because the *RLC* circuit is strongly coupled ($\beta_L = 10$) to the Josephson junction, that contains reactive elements.

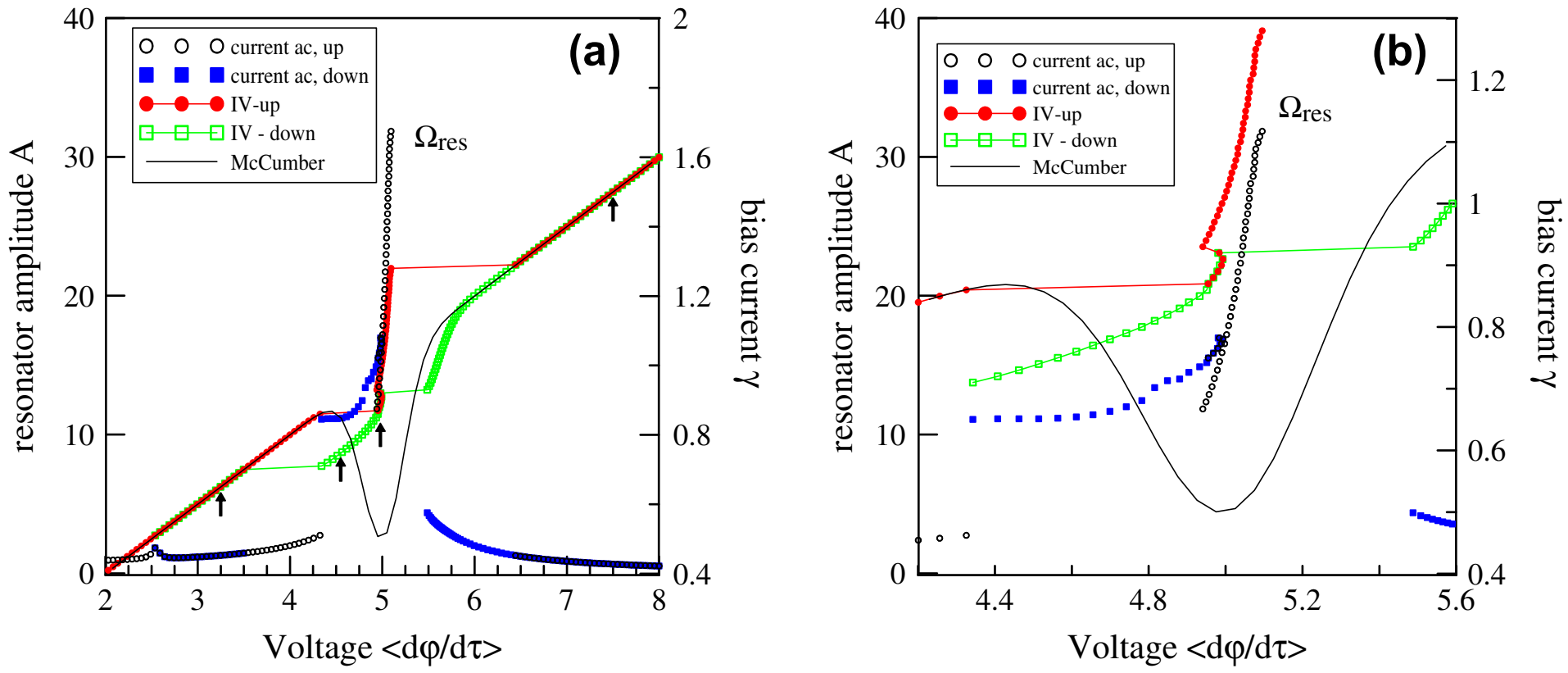

**Fig. 2.** The IV curve and the current in the *RLC* resonator. The bias current is swept from low and high values (a) and an enlargement of the central area where the backbending analogous to the phenomenon observed in stacks of BSCCO [8]. The parameters of the simulations are: $\alpha_0 = 0.2$, $\alpha_1 = 0.5$, $\Omega_{NDR} = 5$, $\Delta = 0.1$, $\beta_L = 10$, $Q = 100$, $\Omega_r = 4$.

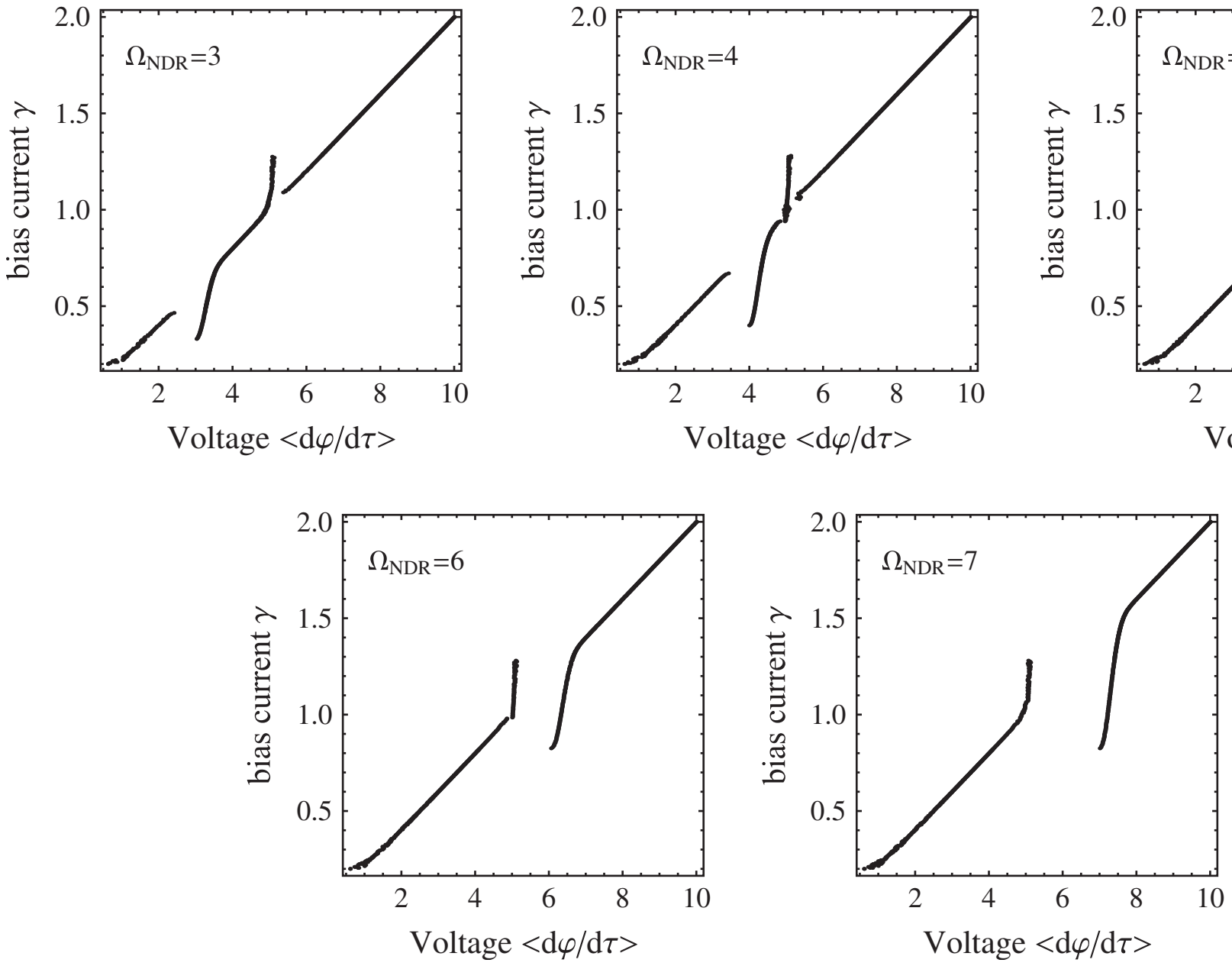

**Fig. 4.** IV curves for different values of the center $\Omega_{NDR}$ of the NDR shunt. Parameters of the simulations are: $\alpha_0 = 0.2,\ \alpha_1 = 0.5,\ \Delta_1 = 0.1,\ \beta_L = 10,\ Q = 100,\ \Omega_r = 4$.

To illustrate the effect, we plot in Fig. 3 the behavior of the peak of Fig. 2, that we call $\Omega_{res}$. From Fig. 3 it is evident that an exponential dependence of the resonance as function of the square root of the coupling. To our knowledge this dependence has not yet been observed in other simulations. We also notice that the exponential behavior rules out some simple scaling of the parameters, and it is probably due to a strong nonlinear interaction between the cavity and the JJ dynamics.

In Figs. 4 and 5 we show the IV curve and the power as a function of the relative position of the cavity resonance ($\Omega_r$) and of the NDR region ($\Omega_{NDR}$). The emitted power is little changed by the NDR, but a "forbidden" gap region appears, in correspondence to the negative differential resistance.

In Fig. 6 we show the time dynamics of the current across the *RLC* circuit. It is evident that in the region of high power the signal is not just a sinusoidal waveform, but a modulated carrier signal. To quantify the non-harmonic component we introduce the moving average $M(\tau)$ of the square of the current in the *RLC* circuit, here $T = 2\pi/V$ is the period of the solution

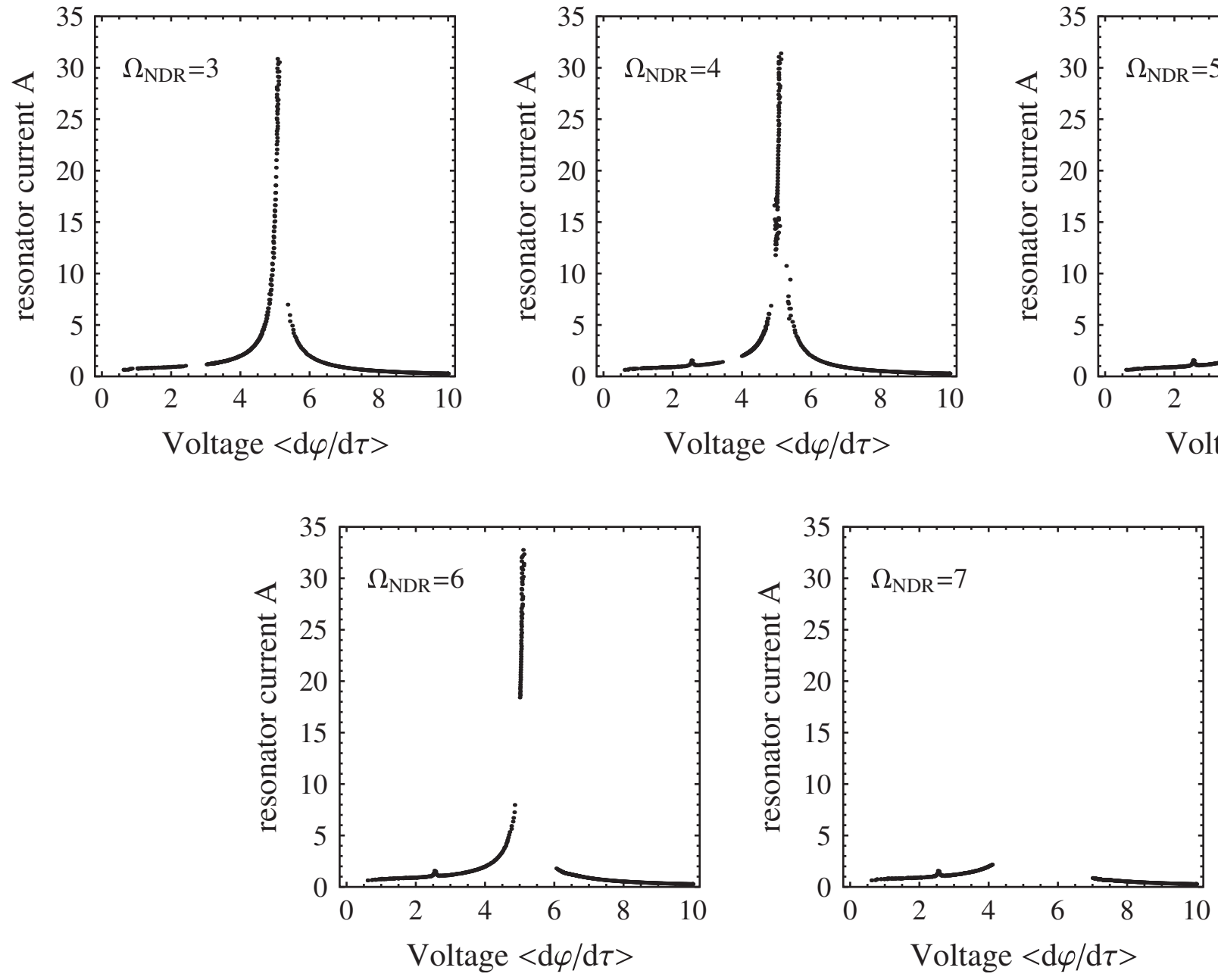

**Fig. 5.** The current through the *RLC* resonator as function of the junction voltage for different values of the center of the NDR shunt. Parameters of the simulations are: $\alpha_0 = 0.2,\ \alpha_1 = 0.5,\ \Delta_1 = 0.1,\ \beta_L = 10,\ Q = 100,\ \Omega_r = 4$.

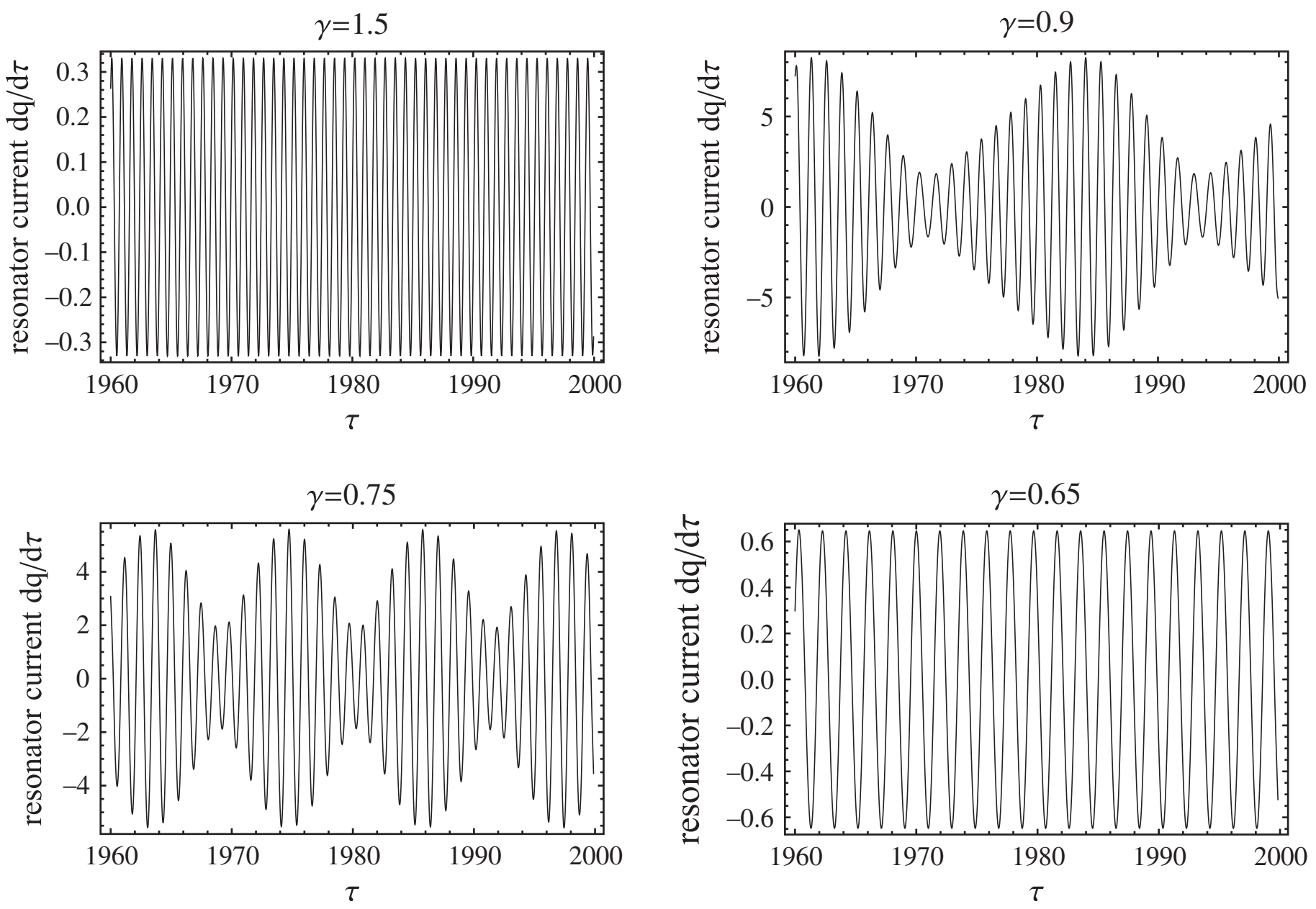

**Fig. 6.** Evolution of the current in the cavity for different values of the bias current $\gamma$, as indicated by the arrows in Fig. 2.

$$M(\tau) = \frac{1}{T}\int_{\tau}^{\tau+T}\left(\frac{dq}{d\tau}\right)^2 d\tau. \tag{8}$$

The variance of $M$ over a long time $\tau_f$ is an estimate of the non-harmonic content:

$$\sigma^2 = \frac{1}{\tau_f}\int_0^{\tau_f}[M(\tau) - \langle M(\tau)\rangle]^2 d\tau. \tag{9}$$

In fact, in Fig. 7 it is clear that such harmonic content increases when the power in the cavity increases. This is expected, for a nonlinear system is likely to generate harmonics when the oscillations are large, however, it indicates that a portion of the available power might be lost exactly where the emission is larger.

Finally, we have also considered the case of $N$ Josephson junctions coupled only through the common cavity. This modifies the equations as follows

$$\frac{d^2\phi_i}{d\tau^2} + \alpha\frac{d\phi_i}{d\tau} + \sin\phi_i + \frac{dq}{d\tau} = \gamma \qquad i = 1,\ldots,N \tag{10}$$

$$\frac{d^2q}{d\tau^2} + \frac{\Omega_r}{Q}\frac{dq}{d\tau} + \Omega_r^2 q - \beta_L\sum_{i=1}^{N}\frac{d\phi_i}{d\tau} = 0. \tag{11}$$

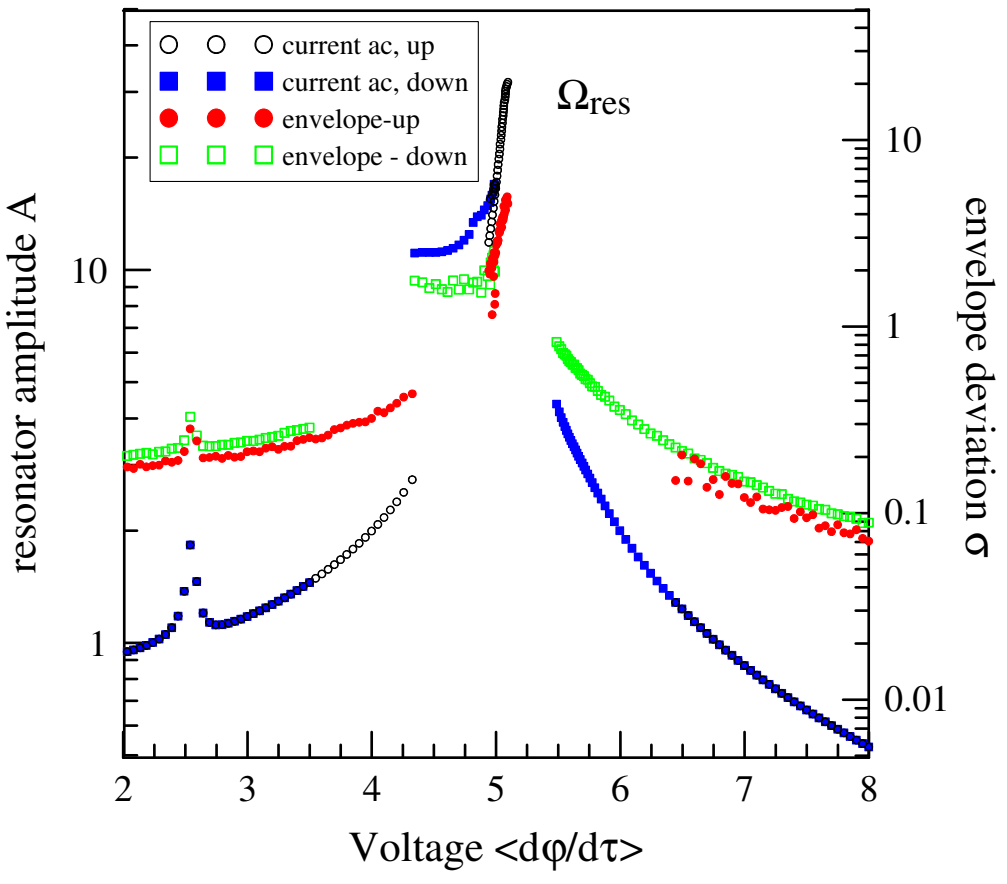

**Fig. 7.** Power and distortions as a function of the voltage. Parameters of the simulations are the same as in Fig. 2.

The qualitative picture is left unchanged in comparison to the previous results and also in this case the coupling $\beta_L$ affects the resonance. We notice that a linear conductance is employed in Eq. (10). In fact, the nonlinear conductance mimics the behavior of a stack of JJ [13], and therefore it is appropriate to employ just a single equation for the whole BSCCO crystal.

## 4. Conclusions

We have found that the interaction between arrays of JJ and a linear resonator gives rise to a resonant frequency at *higher* frequency with respect to the pure resonance of the *RLC* circuit. The increase in the frequency depends in a nontrivial manner upon the coupling. The phenomenon is left unchanged by the presence of a nonlinear conductance in the JJ, or by the insertion of several junctions. The harmonic content of the signal is not uniform across the parameter space when the conditions are such that emission is at a maximum, also the harmonic distortion increases. The nonlinear conductance introduces hysteresis, and therefore the bias scheme is essential to reach the region of high power emission.

## Acknowledgements

We thank COST program MP1201 Nanoscale Superconductivity: Novel Functionalities through Optimized Confinement of Condensate and Fields (NanoSC-COST) for partial support. GF's work at the DTU was supported by the COST "Short Term Scientific Mission" COST-STSM-ECOST-STSM-MP1201-291013-036429. VP thanks INFN (Italy) for partial support.

## References

[1] L. Ozyuzer, A.E. Koshelev, C. Kurter, N. Gopalsami, Q. Li, M. Tachiki, K. Kadowaki, T. Yamamoto, H. Minami, H. Yamaguchi, T. Tachiki, K.E. Gray, W.-K. Kwok, U. Welp, Emission of coherent THz radiation from superconductors, Science 318 (2007) 1291–1293.

[2] H.B. Wang, S. Guenon, B. Gross, J. Yuan, Z.G. Jiang, Y.Y. Zhong, M. Grunzweig, A. Iishi, P.H. Wu, T. Hatano, D. Koelle, R. Kleiner, Coherent terahertz emission of intrinsic Josephson junction stacks in the hot spot regime, Phys. Rev. Lett. 105 (2010) 057002.

[3] U. Welp, K. Kadowaki, R. Kleiner, Superconducting emitters of THz radiation, Nat. Photonics 7 (2013) 702–710.

[4] G.R. Berdiyorov, S.E. Savel'ev, M.V. Milošević, F.V. Kusmartsev, F.M. Peeters, Synchronized dynamics of Josephson vortices in artificial stacks of SNS Josephson junctions under both dc and ac bias currents, Phys. Rev. B 87 (2013) 184510.
[5] P. Addesso, G. Filatrella, V. Pierro, Characterization of escape times of Josephson junctions for signal detection, Phys. Rev. E85 (2012) 016708.
[6] A. Barone, G. Paternó, Physics and Applications of the Josephson Effect, John Wiley & Sons, New York, 1982.
[7] H. Asai, M. Tachiki, K. Kadowaki, Proposal of terahertz patch antenna fed by intrinsic Josephson junctions, Appl. Phys. Lett. 101 (2012) 112602.
[8] K. Kadowaki, H. Yamaguchia, K. Kawamata, T. Yamamoto, H. Minami, I. Kakeya, U. Welp, L. Ozyuzer, A. Koshelev, C. Kurter, K.E. Gray, W.-K. Kwok, Direct observation of tetrahertz electromagnetic waves emitted from intrinsic Josephson junctions in single crystalline $Bi_2Sr_2CaCu_2O_{8+\delta}$, Physica C 486 (2008) 634–639.
[9] S. Lin, A.E. Koshelev, Linewidth of the electromagnetic radiation from Josephson junctions near cavity resonances, Phys. Rev. B 87 (2013) 214511.
[10] P. Barbara, A.B. Cawthorne, S.V. Shitov, C.J. Lobb, Stimulated emission and amplification in Josephson junction arrays, Phys. Rev. Lett. 82 (1999) 1963–1966.
[11] G. Filatrella, N.F. Pedersen, C.J. Lobb, P. Barbara, Synchronization of underdamped Josephson-junction arrays, Eur. Phys. J. B 34 (2003) 3–8.
[12] S. Madsen, N.F. Pedersen, P.L. Christiansen, Repulsive fluxons in a stack of Josephson junctions perturbed by a cavity, Physica C 468 (2008) 649–653.
[13] S. Madsen, N.F. Pedersen, P.L. Christiansen, Stacked Josephson junctions, Physica C 470 (2010) 822–826.
[14] G. Filatrella, N.F. Pedersen, The mechanism of synchronization of Josephson arrays coupled to a cavity, Physica C 372–376 (2002) 11–13.
[15] M. Tachiki, K. Ivanovic, K. Kadowaki, T. Koyama, Emission of terahertz electromagnetic waves from intrinsic Josephson junction arrays embedded in resonance LCR circuits, Phys. Rev. B 83 (2011) 014508.
[16] G. Filatrella, V. Pierro, N.F. Pedersen, M.P. Sørensen, Negative Differential Resistance due to Nonlinearities in Single and Stacked Josephson Junctions, Trans on Appl. Superc., 2013, http://dx.doi.org/10.1109/TASC.2014.2311383.